# Endoscopy Classification Model Using Swin Transformer and Saliency Map

*Zahra Sobhaninia[1], Nasrin Abharian[1], Nader Karimi[1], Shahram Shirani[2], Shadrokh Samavi[1,2,3]*

[1]Department of Electrical and Computer Engineering, Isfahan University of Technology, 84156-83111, Iran
[2]Department of Electrical and Computer Engineering, McMaster University, L8S 4L8, Canada,
[3]Computer Science Department, Seattle University, Seattle 98122 USA

## ABSTRACT

*Endoscopy is a valuable tool for the early diagnosis of colon cancer. However, it requires the expertise of endoscopists and is a time-consuming process. In this work, we propose a new multi-label classification method, which considers two aspects of learning approaches (local and global views) for endoscopic image classification. The model consists of a Swin transformer branch and a modified VGG16 model as a CNN branch. To help the learning process of the CNN branch, the model employs saliency maps and endoscopy images and concatenates them. The results demonstrate that this method performed well for endoscopic medical images by utilizing local and global features of the images. Furthermore, quantitative evaluations prove the proposed method's superiority over state-of-the-art works.*

## 1. INTRODUCTION

In the last decade, the attention of machine learning researchers has been drawn to automatic diagnosis in medical fields. Gastrointestinal (GI) disorders detection is one of these issues worth addressing as diagnosing the GI disease in the early stages can guarantee complete recovery. For instance, early recognition of colon cancer, one of the three most common cancers [1], could considerably affect its prognosis and clinical management [2]. Although endoscopy is one of the most significant tests currently available for screening for GI disorders, it is a challenging procedure that requires a substantial amount of time and expertise to be conducted. In addition, extensive diversity in polyp types, patients without early symptoms, and different understandings by specialists have made the endoscopy image classification task more valuable [3]. As shown in Figure 1, the endoscopic images contain different types of lesions that can be normal or abnormal. Accurate classification can help diagnose the type of these lesions and accelerate the treatment plans.

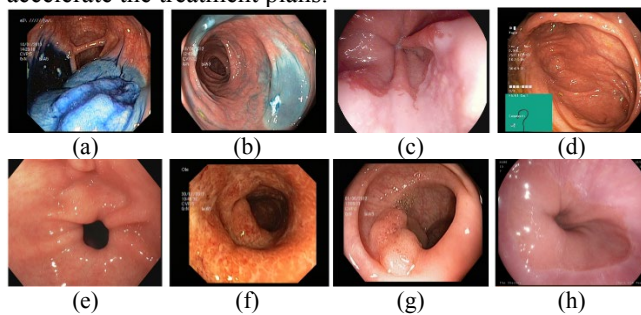

Figure 1: Samples of images in each class of the Kvasir-v2 **[3]** (a): Dyed Lifted Polyps, (b): Dyed Resection Margins, (c): Esophagitis, (d): Cecum, (e): Pylorus, (f): Ulcerative-Colitis, (h): Z-line

Machine learning methods could provide cost-effective strategies for GI diseases because of their accuracy in classification and ability to reduce the diagnosis time. In recent years, deep learning has been developed and has been applied widely in various fields of medical image analysis, such as early diagnosis of anomalies [4], classification [5], and segmentation [6]. Previous works on endoscopy image classification have suggested several methods based on a deep learning approach. For instance, Inés et al. [7] applied a semi-supervised learning technique to achieve fast and light image classification on deep compact architectures.

Raju et al. [8] used a five-stage system that starts from the preprocessing stage and spatial feature extraction using convolutional neural networks (CNNs) and fusion CNN models. In the work of Djenouri et al. [9], different deep learning architectures such as VGG16, ResNet, and DenseNet have been applied, and eventually, the results were determined with ensemble learning. Finally, Gamage et al. [10] proposed an ensemble model of deep features as a single feature vector by combining different pre-trained deep networks, such as DenseNet and ResNet, to classify endoscopy images.

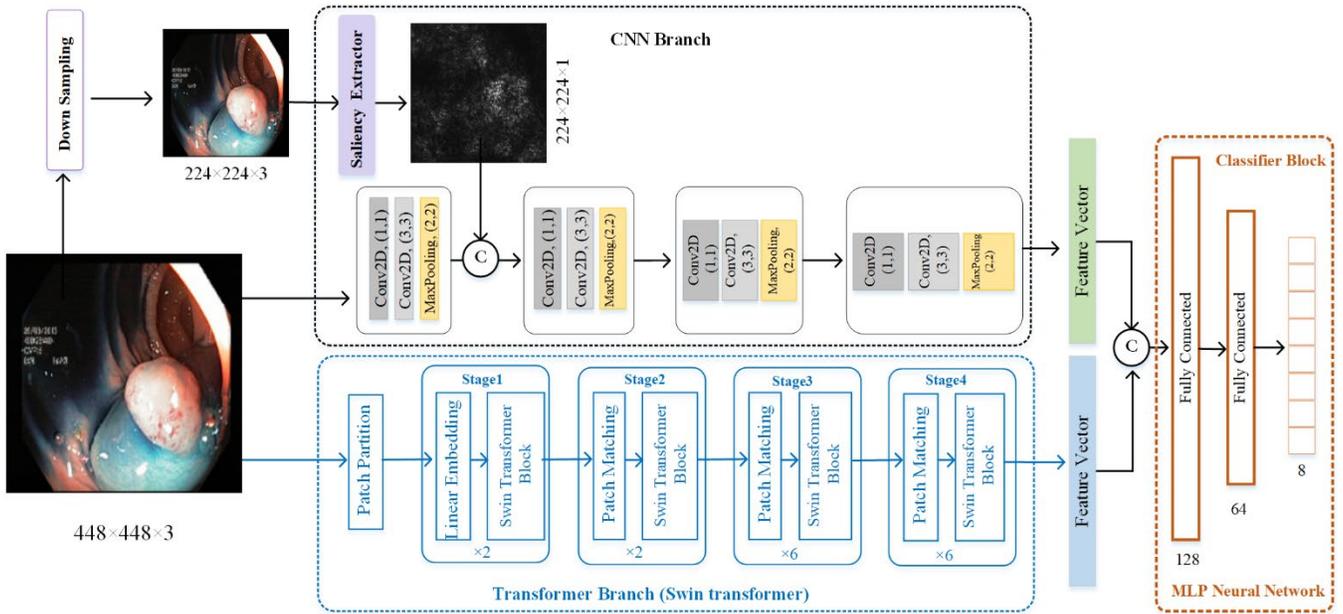

Figure 2: Overview of the SVS-EC model, which contains three components, CNN branch, transformer branch, and classifier block

Afriyie et al. [11] proposed a technique to reduce image noises and improve the CapsNet feature learning process. Ayidzoe et al. [12] focus on eliminating the obstacles that might reduce practitioners' trust in using deep learning models for critical tasks by visualization to overcome challenges of convolutional neural networks such as invariance.

Lately, researchers have studied a novel approach: the application of transformers designed initially to solve natural language processing (NLP) tasks [13]. Transformer models are used to solve the problem of transduction or transformation of input sequences into output sequences in deep learning applications. Significant breakthroughs in applying transformer models in NLP have encouraged researchers to focus on using these models in computer vision tasks, which led to the proposal of a vision transformer (ViT) [14] for functions such as object detection, segmentation, and classification. This new approach has been used for endoscopy image analysis as well. For instance, Bai et al. [15] used a transformer neural network with a spatial pooling configuration for endoscopy image classification. In [16], an approach based on ViT and the transfer learning approach has been proposed for GI diseases. Oh et al. [17] proposed a model based on a multiscale self-attention mechanism that applies 2-stage (ViT) networks for multi-classes endoscopy of gastric cancer.

This paper proposes a model to classify GI disorders, combining two vision-based substantial approaches, a convolutional neural network and a transformer model. The proposed model, SVS-EC (Swin -VGG-Saliency model for multi-label endoscopy classification), is based on using the Swin transformer [18] as a transformer branch and the modified VGG16 model as a CNN branch. The experimental results show that the impact of applying CNN and transformer models jointly instead of using one of them leads to a remarkable improvement in the learning process and an increase in classification criteria.

The rest of this paper is structured as pursues: Section 2 explains the proposed approach in detail. Then, section 3 discusses the experimental results and compares them with state-of-the-art methods. Eventually, in Section 4, the conclusion of the proposed approach is elaborated.

## 2. PROPOSED METHOD

Since the CNNs process the input images locally and layer by layer, they depend on local features principally. However, transformers can better generalize features and process input images globally than CNNs based on their structures. Hence, combining these two views of feature extraction (local and global view) approaches can benefit the model from both learning strategies. Furthermore, in some medical image analyses, one of the important issues is to pay attention not only to the lesions but also to the surrounding regions and the background, which affect the diagnosis of the type of disease. In this regard, constructing a model that contains a CNN network as the local feature extraction process and a transformer model for analyzing images in a global view and performing classification based on these two features provides a better learning process due to considering these two aspects jointly.

Based on this strategy, the proposed method, the SVS-EC model, contains three main components. The first part is a Swin transformer that is for global learning. The second part is a modified VGG16 network as a CNN branch for local learning. The third part is a classifier block. The saliency map is added as a concatenation layer to the VGG pipeline to enhance the VGG16's performance. Figure 2 illustrates the block diagram of the SVS-EC model. After the inputs pass through these two branches, a classification block is used to perform classification based on multi-layer perceptron (MLP) neural networks. Details of these branches are explained in sections A and B.

### 2.1. Transformer branch

In the proposed model, the transformer branch is based on the Swin Transformer [18], one of the new vision transformer models that construct a hierarchical feature representation. In addition, its computational complexity is linearly proportional to the input image size. In this model, unlike the common ViT model, where images are initially divided into patches with a fixed size, the first layer starts with small patches. Next, these paths are merged into bigger ones in the deeper layers. The original Swin transformer splits an image into 4 × 4 patches. Each patch is a colored image with three channels. Thus, a patch has a total of 48 feature-dimensionality.

The performance of this model is associated with the shifted-window-based self-attention method deployed. However, in this work, to emphasize the global learning aspect, the size of the primary patch has been set to 48×48 instead of 4×4. These non-overlapping patches comprise the whole image cited as a "token." A token is the smallest processing unit in the model. When this procedure is done (splitting and rearranging), the transformer model embeds every patch and uses encoder blocks for more feature extraction.

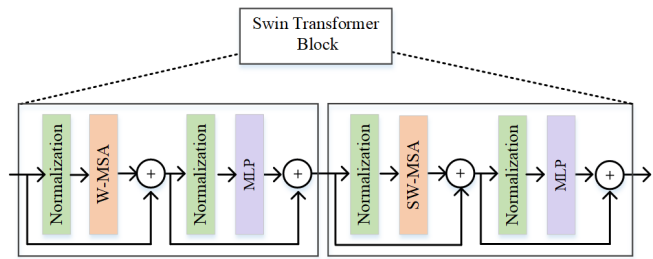

Figure 3: The structure of the Swin Transformer block

As presented in Figure 2, the Swin transformer employed in this study includes two components, patch embedding, and an encoder block. After patch embedding, the input image passes through an encoder layer that extracts the features of the inputs. The Swin transformer encoder block comprises a normalization operation, shifted window multi-head, self-attention module (SW-MSA), and window multi-head self-attention (W-MSA) module instead of standard multi-head self-attention (MSA) module. In the final step, extracted features are enriched with a global average-pooling layer to the dense layer with an MLP of size 64. The Swin encoder architecture is summarized in Figure 3.

### 2.2. CNN branch

In this branch, the modified VGG-16 model has been used for exploiting the CNN approach for feature extraction. In addition, the number of layers, filters, size of kernels, and strides have been changed to decrease the computational cost and prevent overfitting. This is presented in Figure 2.

At first, the input endoscopy images are fed in two convolutional layers with the same filter size and different kernels. Then the feature maps pass from the max-pooling layer to a stride of two pixels to reduce the resolution. After this stage, the saliency map of input images is concatenated to the corresponding layer of the VGG network. At each level, this process repeats, and the number of filters is increased to extract more feature maps. Finally, a CNN branch enriches the semantic information in the last layer of the feature map, a flattened layer with a size of 256 and then two dense layers embedded with 128 and 64.

Passing through the convolutional and max-pooling layers leads to missing some information [19]. Hence, concatenating the saliency map tries to compensate for this loss and helps the network learn essential regions better. This is done by concatenating the crucial areas of endoscopy images.

The saliency map of each image has been extracted to locate crucial areas of the images. Saliency map detection is a procedure to index the image's pixels based on the final score of a deep CNN network. Considering the saliency map of images as the input image's fourth channel can improve the CNN approach's performance by helping the network focus on the lesions and uneven parts of endoscopic images. In this work, to acquire the saliency map of endoscopy images, the images first enter the downsample block to change their dimensions (448×448 to 224×224). Using a CNN network (VGG-16 model) trained on the ImageNet dataset is necessary. By feeding the images to the model, the last layer gives a score for each class. Then the gradient is calculated concerning the highest-class score to determine which pixels in the image are more important to emphasize on them.

As depicted in Figure 2, the MLP neural network that is applied in the last step of completing the classification task includes three fully connected layers: a 128 one-dimensional input layer, a 64 one-dimensional hidden layer, and an outcome layer, which its size is equal to the number of classes. Finally, the outputs of the two branches are concatenated together and fed into the MLP for classification to perform multi-label end-to-end classification.

### 3. EXPERIMENTS AND RESULTS

First, this section explains the dataset, metrics, and results. Then, the impact of each branch is investigated, and finally, to show the superiority of the proposed model, a comparison is made with state-of-the-art models.

## 3.1. Dataset

In this work, the Kvasir-v2 dataset has been used, which contains 8,000 endoscopic images taken from inside the gastrointestinal (GI) tract and has been annotated by medical experts. Eight classes are classified based on lesion removal, pathology, and anatomy. The resolution of images varies from 720×576 up to 1920×1072 pixels. In some cases, a green picture is embedded in the images to support the interpretation of the image by illustrating where the endoscope is located in the bowl. Data was collected at Vestre Viken Health Trust in Norway [3]. The dataset is divided into three sets for training and evaluating the proposed method: a training set with 4800 images, a validation set with 2000 images, and a test set with 1200 images.

## 3.2. Metrics

Criteria used for evaluating the proposed model are the standard computer vision metrics such as the area under the curve (AUC), accuracy, recall, precision, and F1-score. Detailed information about the evaluation metrics can be found in [20].

## 3.3. Results

The quantitative results of using the SVS-EC model on Kvasir-v2 are presented in this section. First, the impact of adding each branch has been investigated, and next, the results of the proposed approach have been compared with state-of-the-art works.

Different experiments have investigated the impact of using a saliency map in the CNN branch and using the transformer and CNN branches jointly. The results of these are shown in Table 1. As shown, concatenating the saliency maps to the corresponding layer of modified VGG16 helps the network's performance. In the following, the Swin transformer model has been used, and finally, the impact of using the modified VGG16 and Swin model jointly as the SVS-EC model has been investigated. As shown, the performance of the SVS-EC model, which contains two local and global learning, performs by significant improvement.

Table 1: Investigation of the existence of each branch in the proposed model on quantitative results

| Method | F1-score | Accuracy | AUC | Recall | Precision |
|---|---|---|---|---|---|
| Modified VGG16 | 81.70 | 83.45 | 0.933 | 81.04 | 82.39 |
| Modified VGG 16 + Saliency Map | 83.38 | 86.80 | 0.96 | 82.87 | 83.91 |
| Swin Transformer | 90.75 | 92.89 | 0.98 | 90.06 | 91.47 |
| SVSMLC | 93.45 | 96.88 | 0.991 | 92.68 | 94.23 |

The quantitative analysis from the state-of-the-art models is summarized in Table 2. As can be seen, Inés's [7] method achieves an F1-score of 89% by using the ResNet-50 model. Afriyie [11] applied the denoising technique in its proposed CapsNet framework and obtained 94.16 for accuracy. The five-stage system proposed by Raju et al. [8] achieved an AUC of 0.989. The proposed method results show that considering two aspects of learning, local and global learning, leads to significant improvement in results and provides a good fit in the classification task of Kvasir-v2 dataset images, as it reported SVSMLC gained the best results in comparison to other works.

Table 2: Quantitative results of state-of-the-art methods that have used the Kvasir-v2 dataset.

| Method | F1-score | Accuracy | AUC | Recall | Precision |
|---|---|---|---|---|---|
| [7] | 89 | - | - | - | - |
| [8] | - | 84 | 0.989 | - | - |
| [9] | 77 | 79 | - | - | - |
| [11] | 86.6 | 94.16 | - | - | 83.1 |
| [12] | - | 80.57 | 0.812 | - | - |
| **SVSMLC** | **93.45** | **96.88** | **0.991** | **92.68** | **94.23** |

## 4. CONCLUSION

In this paper, a model has been proposed that is based on local and global learning strategies by utilizing the combination of two feature extraction branches, the Swin transformer, and modified VGG16. First, the local features are extracted and emphasized in CNN networks by concatenating with a saliency map. Then, by applying transformer models, the images are analyzed globally. Finally, these two aspects join together to provide better performance for the classification process. The experiments illustrated that our proposed method could achieve notable results on the Kvasir-v2 dataset, and it has a superior performance in endoscopic image classification. The presented method can be extended to boost the performance of medical image classification and be applied to various types of medical imaging.